\documentclass[intlimits,twoside,a4paper]{article}
\usepackage[cp1251]{inputenc}

\def\be{\begin{equation}}
\def\ee{\end{equation}}

\usepackage[eqsecnum]{cmpj3}

\issue{2021}{24}{3}{33604}
\doinumber{10.5488/CMP.24.33604}

\title[Solvation of nonpolar solute]%
{Pressure dependence of solvation of non-polar solute in simple model of water%
\thanks{Dedicated to Professor Yurii Kalyuzhnyi on the occasion of his 70$^\mathrm{th}$ birthday.}}
\author[T. Urbic]{T. Urbic\orcid{0000-0001-6440-0220}}
\address{Faculty of Chemistry and Chemical Technology,
University of Ljubljana, Vecna Pot 113, SI-1000 Ljubljana, Slovenia}


\date{Received May 03, 2021, in final form July 21, 2021}

\begin{document}

\maketitle

\begin{abstract}
We modelled the aqueous solvation of a nonpolar solute as a function of the radius, temperature and pressure. In this study a simple two-dimensional Mercedes-Benz (MB) water model was used in NPT Monte Carlo simulations. This model has previously been shown to qualitatively predict the volume anomalies of pure water and the free energy, enthalpy, entropy, heat capacity, and volume change in order to insert a nonpolar solute into water. Here, we extended the studies of solvation of nonpolar solute to examine the pressure dependence and broader range of temperature and size dependence. The model shows two different mechanisms, one for the solvation of large nonpolar solutes bigger than water and the second for smaller solutes. 
%

%
\printkeywords
\end{abstract}

\section{Introduction}

Understanding solute-solvent interactions in aqueous solutions is important in biology and chemistry. A lot of works and several important review papers have been published on this subject~\cite{EK,Franks,sti,Tanford,rob,Pratt,Widom,Griffith,chand,pn,water}.
Various water models with different degrees of sophistication have been
developed to capture the anomalous properties of water and to study
aqueous solutions (for review see~\cite{nezbeda1,Smith}). 
In spite of huge time spent studying the water and its solvation properties, the water is considered to be a poorly understood liquid~\cite{dillreview}. The most important anomalous properties of water are maximum in density, negative expansion coefficient, low compressibility, high and almost constant heat capacity~\cite{water}. The principal challenge in modelling the physics of water comes from the orientation-dependent formation of hydrogen bonds. 

Many properties of the  water and aqueous solutions can be captured by simple models that lack atomic details~\cite{trusket1,trusket2,fennell}. One of the simplest model for water is the Mercedes-Benz (MB) model~\cite{mbs} which was originally proposed by Ben-Naim in 1971~\cite{ben,ben2}. Each molecule is described as disk with Lennard-Jones (LJ) interactions and three radial arms arranged as in the
MB logo to mimic the formation of hydrogen bonds (HB). This HB interaction is an orientation-dependent interaction. 

Simplified models are of interest
due to the insights they offer that are not obtainable from all-atom computer simulations. Simpler models are more flexible in providing insights and illuminating concepts, and do not require big computer resources. Second, simple models can explore a much broader range of conditions and external variables. Meanwhile, simulating a detailed model may predict the behaviour at a single temperature and pressure, and a simpler model can be used to study a whole phase diagram of temperatures and pressures much faster. Third, analytical models can provide functional relationships for engineering applications and lead to improved models of greater computational efficiency. Fourth, simple models can be used as a polygon to develop and study theoretical methods. Our interest in using the MB model is that it serves as one of the simplest models of an orientationally dependent liquid. Thus, it can serve as a testbed for developing analytical theories that
might ultimately be useful for more realistic models. Another advantage of the MB model, compared to more realistic water models, is that the underlying physical principles can be more readily explored and visualised in two dimensions. 

For the MB model, NPT Monte Carlo simulations have shown that the MB model qualitatively predicts the density anomaly, the minimum in isothermal compressibility as a function of temperature, a large heat capacity, as well as the experimental trends for thermodynamic properties of solvation of nonpolar solutes~\cite{mbs,anda,mbs2,noel,silver,hribar,dias1} and cold denaturation of proteins~\cite{dias2}. The 2D MB model was also extended to 3D by Bizjak et al.~\cite{bizjak1,bizjak2} and Dias et al.~\cite{dias1,dias3} and was studied using computer simulations~\cite{bizjak1,bizjak2,dias1,dias2}. The 2D model was also extensively studied using analytical methods such as integral equation and thermodynamic perturbation theory~\cite{urbic1,urbic2,urbic3,urbic4,urbic5,urbic6,urbic7} and statistical mechanic modelling~\cite{urbic,urbic10,urbic11}. Recently, phase diagram of liquid part and percolation curve of the model was also calculated and reported~\cite{urbic9}. The MB model has also been used to study the systems with water molecules confined in partially quenched disordered matrix~\cite{urbic4, kurtjak2014, kurtjak2015} and within small geometric spaces~\cite{urbic7, urbic2006}. Non-equilibrium Monte Carlo and molecular dynamics simulations were used to study the effect of translational and rotational degrees of freedom on the structural and thermodynamic properties of the simple Mercedes-Benz water model~\cite{um1}. By holding one of the temperatures constant and varying the other one, the effect of faster motion in the corresponding degrees of freedom on the properties of the simple water model was investigated. For this case, we also developed an analytical theory for studying rotational and translational degrees of freedom, by applying integral equation theory and thermodynamic perturbation theory~\cite{ogrin1,ogrin2}. In this work we extended the studies of solvation of nonpolar solute to study the effect of size, pressure and temperature as well as the depth of LJ potential between water and solute.

\section{Model}

\begin{figure}[!t]
  \centerline{\includegraphics[width=80mm]{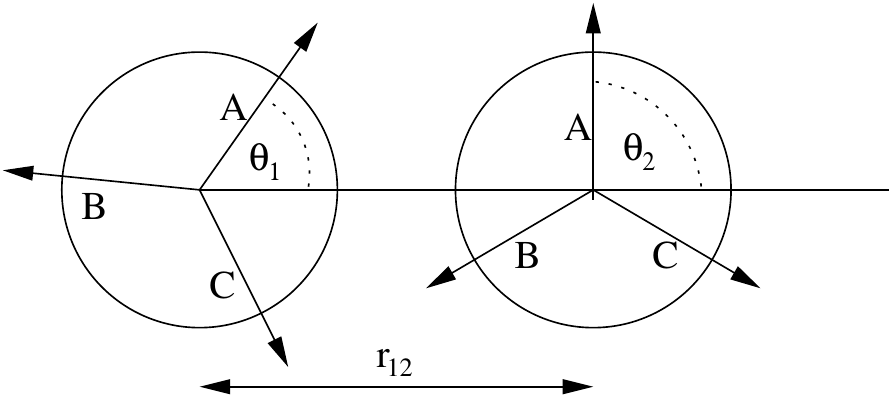}}
  \caption{The MB particles.}
  \label{fig:sh.pot}
\end{figure}
In the MB model, the water molecules are modelled as two-dimensional disks with Lennard-Jones potential and three arms placed so that the angle between them is 120$^\circ$~\cite{ben,mbs} to mimic the formation of hydrogen bonds between molecules. The interaction between two molecules $i$ and $j$ is the sum of Lennard-Jones (LJ) and hydrogen-bonding (HB) term 

 \be
U(\vec{X}_i,\vec{X}_j)=U_{\rm{LJ}}(r_{ij})+U_{\rm{H
B}}(\vec{X}_i,\vec{X}_j) \label{mb1} \ee
and depends on the distance between molecules as well as on their relative orientation ($\vec{X}_i$ is the vector of the position and orientation of the $i$-th particle). The Lennard-Jones part of the potential is calculated in a standard way as follows: 

\be
U_{\rm{LJ}}(r_{ij}) = 4\varepsilon_{\rm{LJ}} \left[ \left( \frac{\sigma_{\rm{LJ}}}{r_{ij}} \right)^{12} -
\left(  \frac{\sigma_{\rm{LJ}}}{r_{ij}} \right)^6 \right]. \label{lj1}
\ee
 $\varepsilon_{\rm{LJ}}$ is the depth and $\sigma_{\rm{LJ}}$ is the contact distance of the LJ potential. The term that describes hydrogen bonds is a sum of all interactions $U_{\rm{HB}}^{kl}$ between the arms $k$ and $l$ of molecules $i$ and $j$, respectively
 
 \be 
 \label{hb1} U_{\rm{H
		B}}(\vec {X}_i,\vec{X}_j)=\sum_{k,l=1}^3 U_{\rm{H
		B}}^{kl}(r_{ij},\theta_i,\theta_j), \ee
where $\theta_i$ is the orientation of $i$-th molecule, interaction between two arms is modelled with Gaussian function depending on the distance between the interacting molecules and their orientation.

\begin{multline} \label{hb3}
U_{\rm{HB}}^{kl}(r_{ij},\theta_i,\theta_j)=\varepsilon_{\rm{H
		B}}
G(r_{ij}-r_{\rm{HB}})
G(\vec{i}_k\vec{u}_{ij}-1)G(\vec{j}_l\vec{u}_{ij}+1)\\ =\varepsilon_{\rm{HB}}
G(r_{ij}-r_{\rm{HB}}) G(\cos(\theta_i+{\scriptstyle
\frac{2\piup}{3}}(k-1))-1) G(\cos(\theta_j+{\scriptstyle
\frac{2\piup}{3}}(l-1))+1). 
\end{multline}
$G(x)$ is an unnormalized Gaussian function: 

\be
G(x)=\exp{\left(-{\frac{x^2}{2\sigma^2}}\right)}, \label{gaus1} \ee
$\varepsilon_{\rm{HB}}$ is the HB energy and $r_{\rm{HB}}$ is a HB distance. $\vec{u}_{ij}$ is the unit vector along $\vec{r}_{ij}$ and
$\vec{i}_k$ is the unit vector representing the $k$-th
arm of the $i$-th particle. When the arms on two molecules are parallel, pointing toward the center of the other molecule and the distance between molecules is $r_{\rm{H
B}}$, the strongest possible hydrogen bond is formed. Units used are the same as in previous studies: energies were expressed in $|\varepsilon_{\rm{H
B}}|$ and lengths in $r_{\rm{H
B}}$. Hence, parameter for hydrogen-bond energy, $\varepsilon_{\rm{HB}}$, equalled $-1$, and hydrogen-bond length equalled $1$. LJ interaction potential parameters were set to: $\varepsilon_{\rm{LJ}} = 0.1|\varepsilon_{\rm{HB}}|$ and $\sigma_{\rm{LJ}} = 0.7 r_{\rm{HB}}$. The width of Gaussian ($\sigma =
0.085 r_{\rm{HB}}$) is small enough, so that a direct hydrogen bond is more
favourable than a bifurcated one.

Hydrophobic solute was modelled as a disk with Lennard-Jones potential. The depth of solute's LJ potential was the same as depth of water's LJ potential for most calculations, while the $\sigma_{\rm{LJ}}$, of solute's LJ potential were varied to represent solutes of various sizes (from 0.1 to 5.00). When the effect of depth of LJ potential was studied, the $\varepsilon_{\rm{LJ}}$ was varied from 0.01 to 0.5. When calculating the interaction between solute and water molecule, Lorentz-Berthelot mixing rules were used~\cite{dakota}.

\begin{figure}[!htb]
	\begin{center}
		(a)\includegraphics[width=62mm]{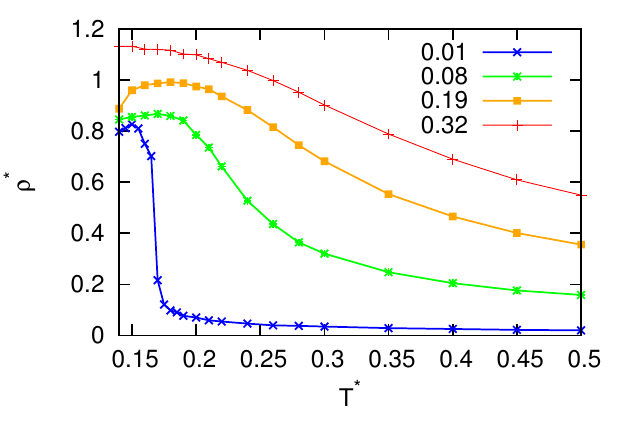}
		(b)\includegraphics[width=62mm]{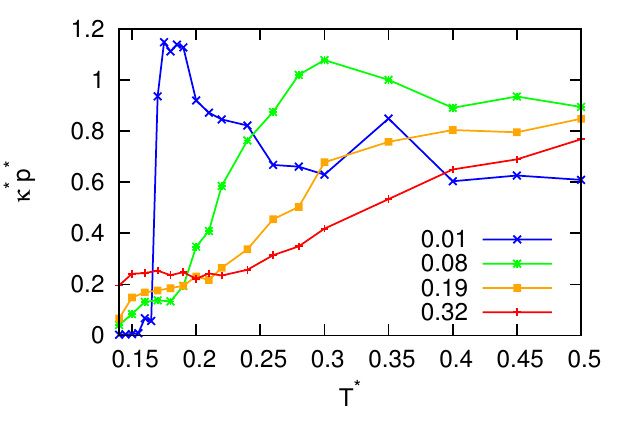}\\
		(c)\includegraphics[width=62mm]{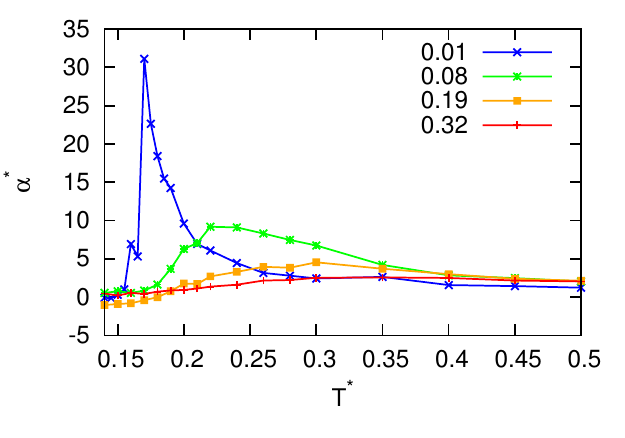}
		(d)\includegraphics[width=62mm]{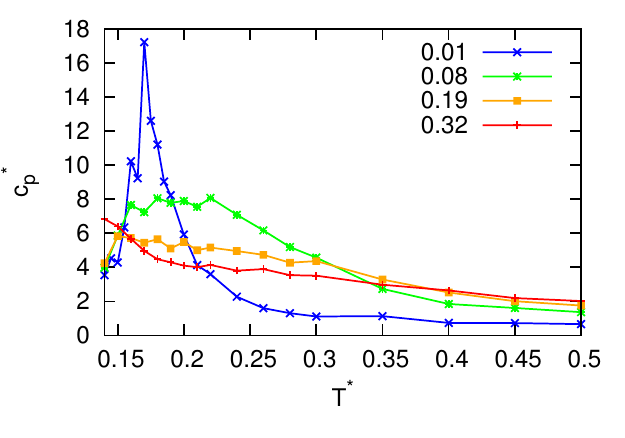}\\
		\caption{(Colour online) Temperature dependence of (a) the density, (b) the isothermal compressibility multiplied by pressure, (c) the thermal expansion coefficient and (d) the heat capacity for various pressures.}
		\label{fig:tt}
	\end{center}
\end{figure}

\section{Monte Carlo simulation details}

\begin{figure}[htb]
\begin{center}
(a)\includegraphics[width=62mm]{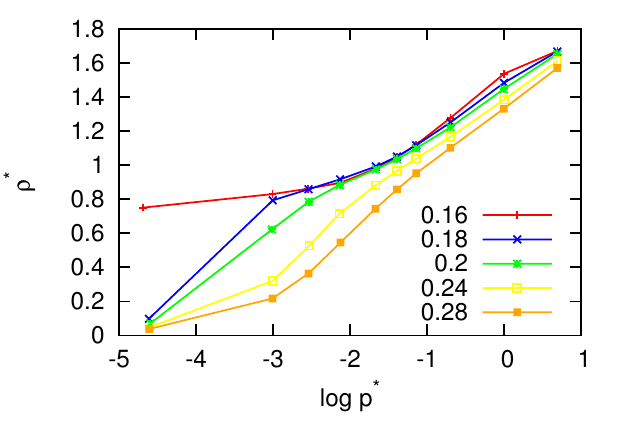}
(b)\includegraphics[width=62mm]{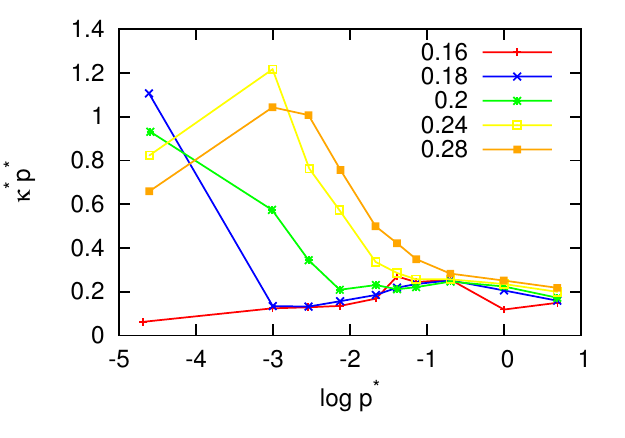}\\
(c)\includegraphics[width=62mm]{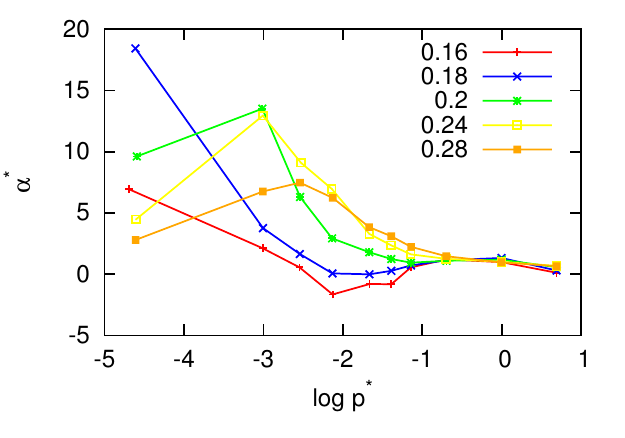}
(d)\includegraphics[width=62mm]{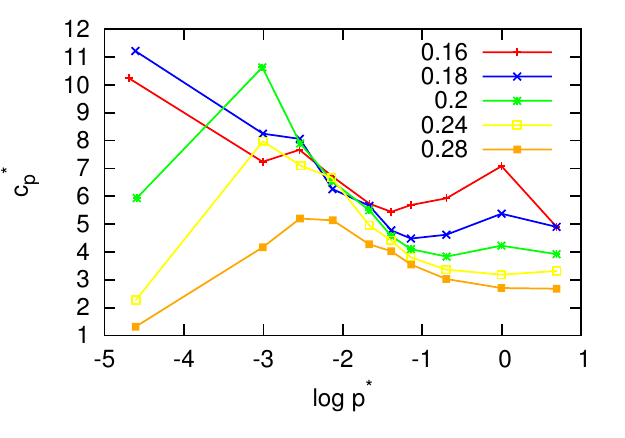}\\
  \caption{(Colour online) Pressure dependence of (a) the density, (b) the isothermal compressibility multiplied by pressure, (c) the thermal expansion coefficient and (d) the heat capacity for various temperatures.}
  \label{fig:pp}
\end{center}
\end{figure}
\begin{figure}[!bht]
	\begin{center}
(a)\includegraphics[width=65mm]{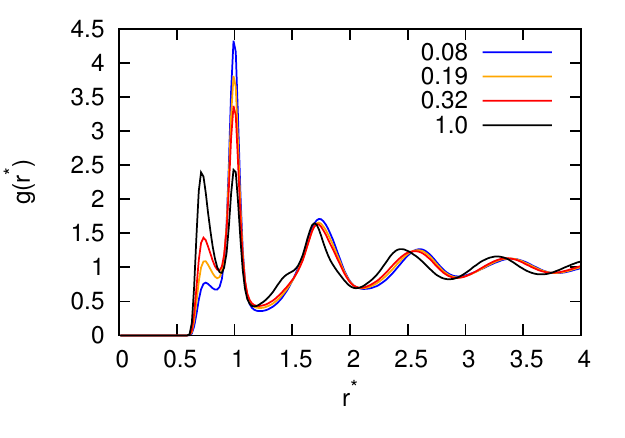}
(b)\includegraphics[width=65mm]{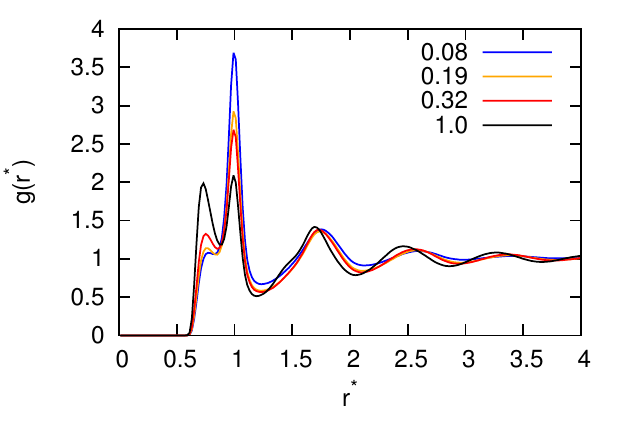}\\
  \caption{(Colour online) Pair correlation functions for various pressures at (a) low temperature $T^*=0.18$, and (b) high temperature $T^*=0.24$.}
  \label{fig:ww}
\end{center}
\end{figure}
\begin{figure}[htb]
	\begin{center}
(a)\includegraphics[width=65mm]{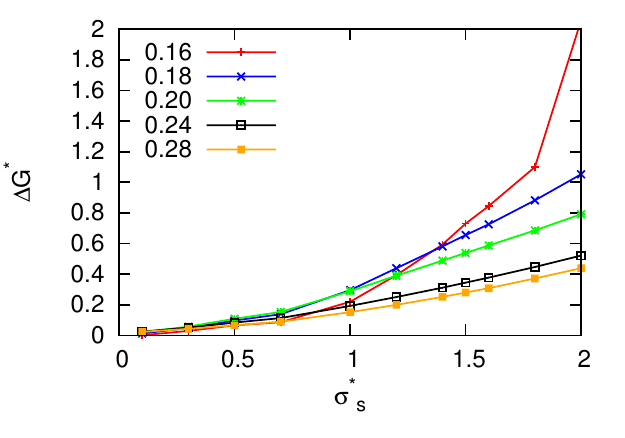}
\includegraphics[width=65mm]{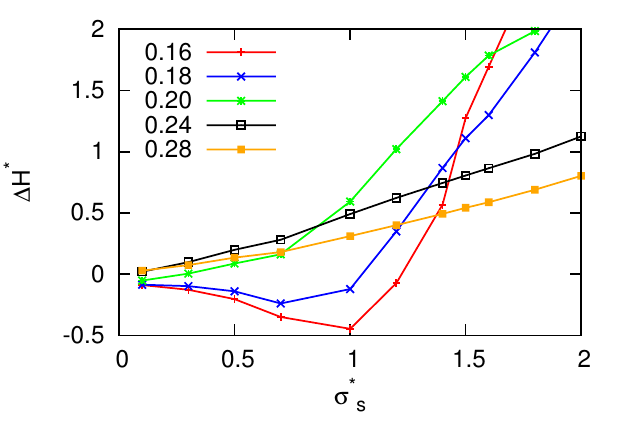}\\
(b)\includegraphics[width=65mm]{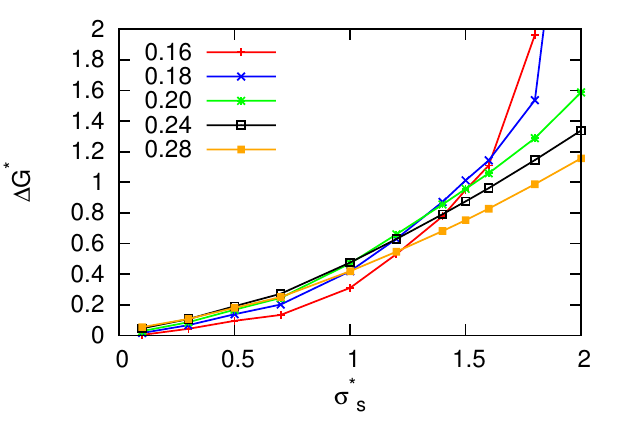}
\includegraphics[width=65mm]{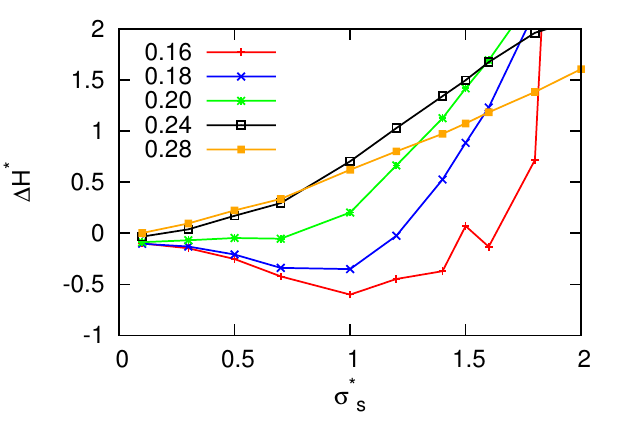}\\
(c)\includegraphics[width=65mm]{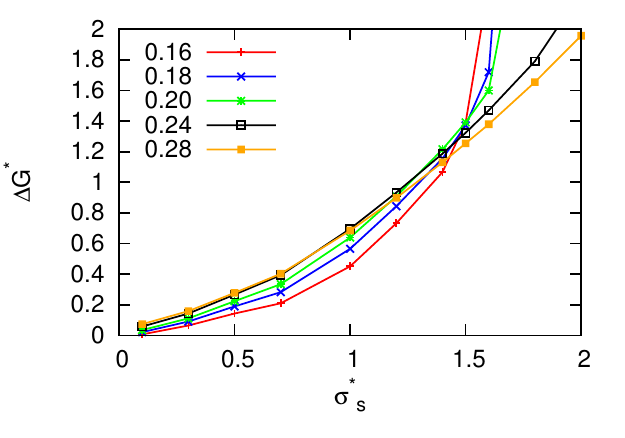}
\includegraphics[width=65mm]{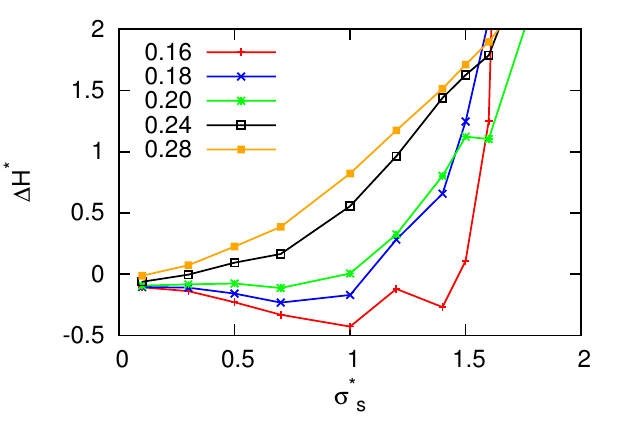}\\
  \caption{(Colour online) Size dependence of transfer free energy and transfer enthalpy of nonpolar solute for pressures (a) $p^*=0.08$, (b)  $p^*=0.19$, and (c) $p^*=0.32$ for various temperatures for depth of solute-solute LJ potential equal to $\varepsilon_{\rm{LJ}}=0.1$ }
  \label{fig:tp2}
\end{center}
\end{figure}
\begin{figure}[htb]
	\begin{center}
(a)\includegraphics[width=65mm]{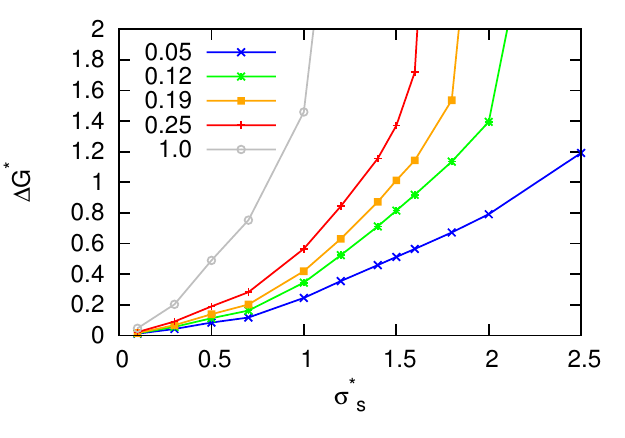}
\includegraphics[width=65mm]{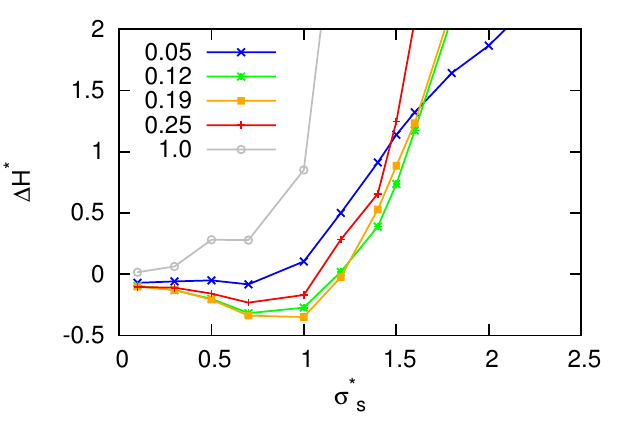}\\
(b)\includegraphics[width=65mm]{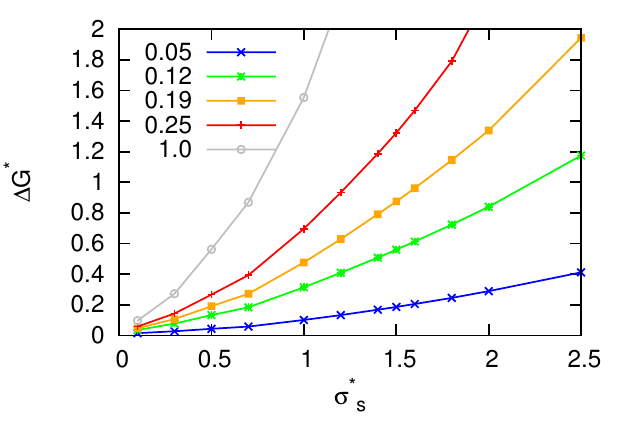}
\includegraphics[width=65mm]{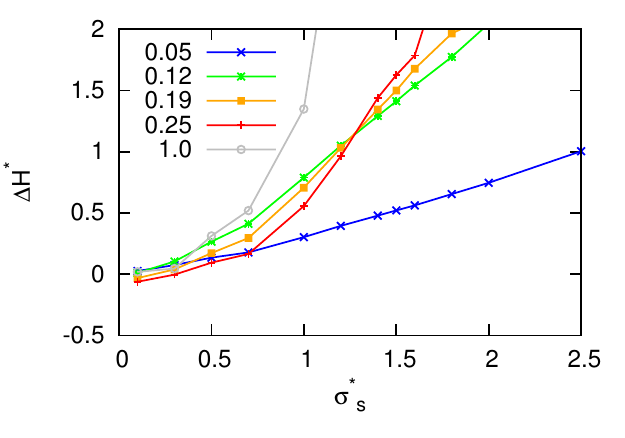}\\
  \caption{(Colour online) Size dependence of transfer free energy and transfer enthalpy of nonpolar solute for temperatures (a) $T^*=0.18$, and (b)  $T^*=0.24$ for various pressures for depth of solute-solute LJ potential equal to $\varepsilon_{\rm{LJ}}=0.1$ }
  \label{fig:tp3}
\end{center}
\end{figure}
\begin{figure}[!htb]
	\begin{center}
(a)\includegraphics[width=65mm]{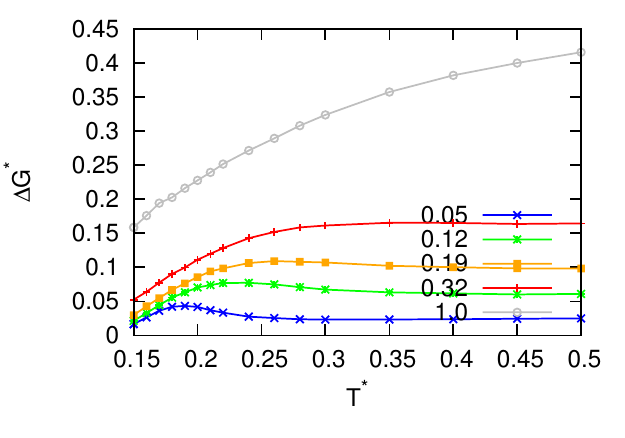}
\includegraphics[width=65mm]{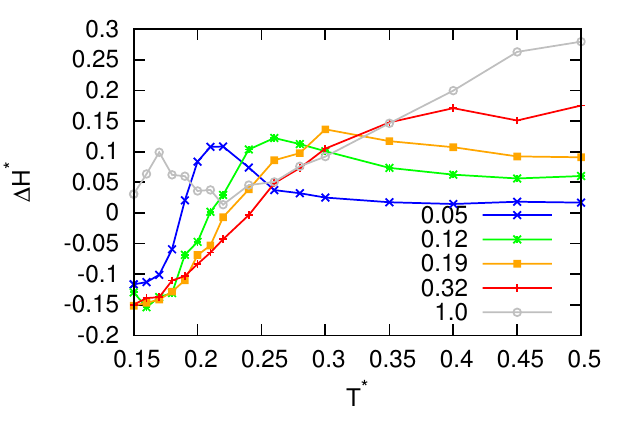}\\
(b)\includegraphics[width=65mm]{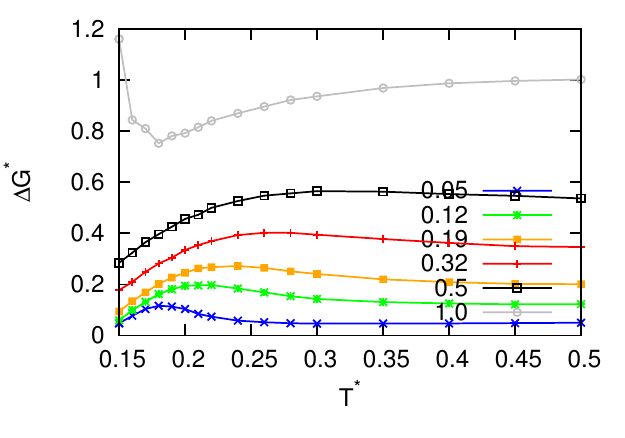}
\includegraphics[width=65mm]{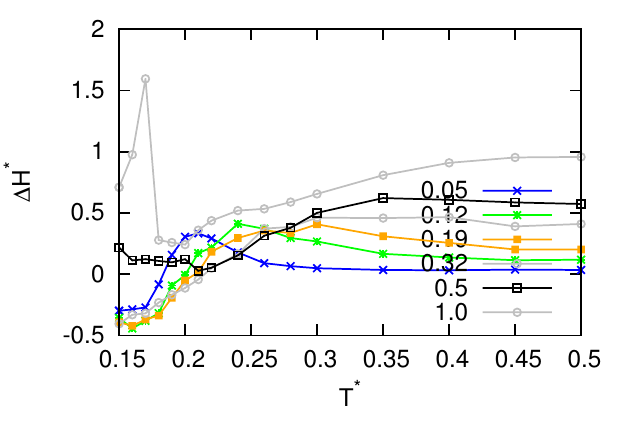}\\
(c)\includegraphics[width=65mm]{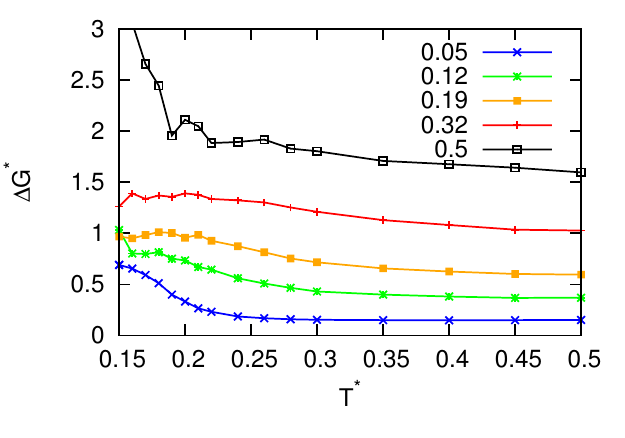}
\includegraphics[width=65mm]{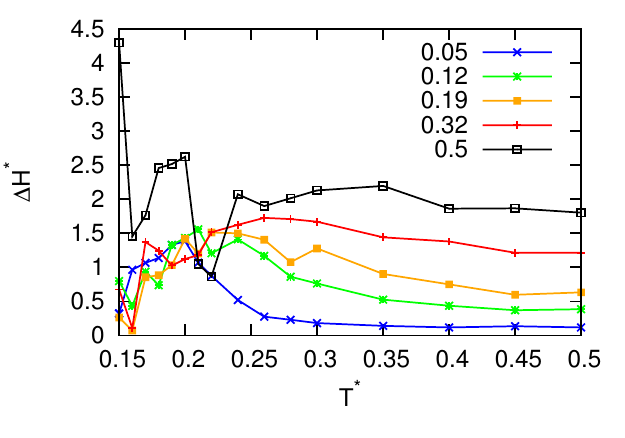}\\
  \caption{(Colour online) Temperature dependence of transfer free energy and transfer enthalpy of nonpolar solute for  sizes of the solution (a) $\sigma_{\rm{LJ}}=0.3$, (b)  $\sigma_{\rm{LJ}}=0.7$, and (c) $\sigma_{\rm{LJ}}=1.5$ for various pressures for depth of solute-solute LJ potential equal to $\varepsilon_{\rm{LJ}}=0.1$ }
  \label{fig:tp1}
\end{center}
\end{figure}
\begin{figure}[!htb]
	\begin{center}
(a)\includegraphics[width=65mm]{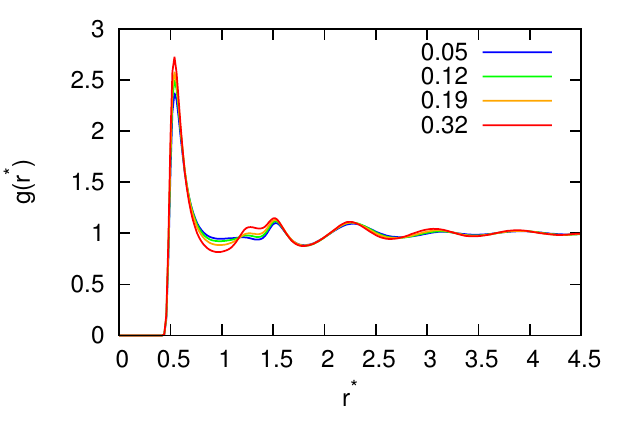}
\includegraphics[width=65mm]{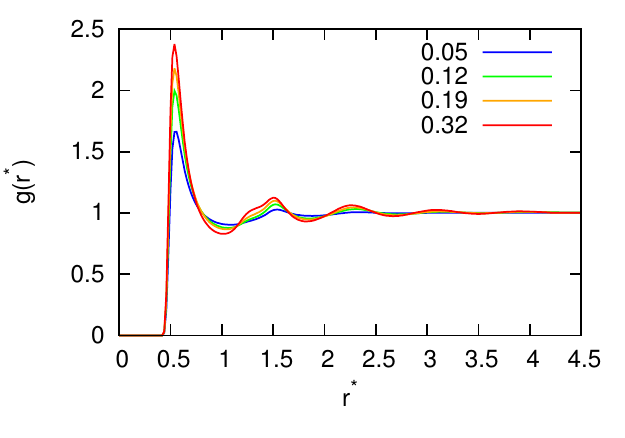}\\
(b)\includegraphics[width=65mm]{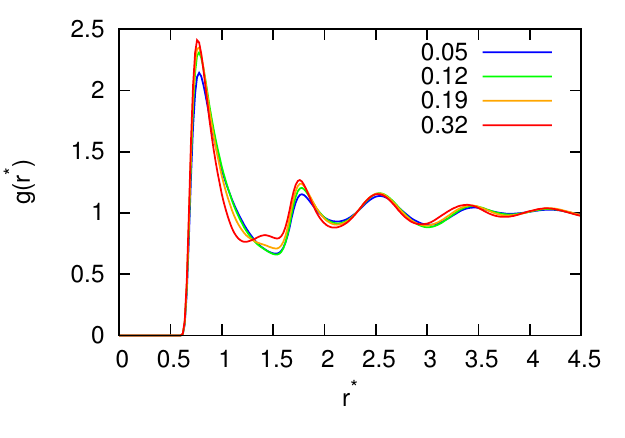}
\includegraphics[width=65mm]{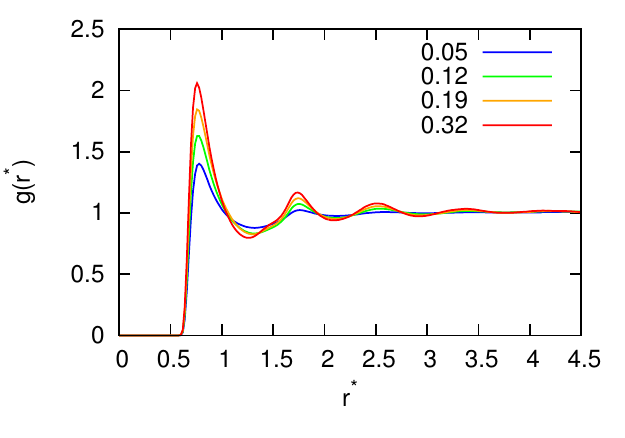}\\
(c)\includegraphics[width=65mm]{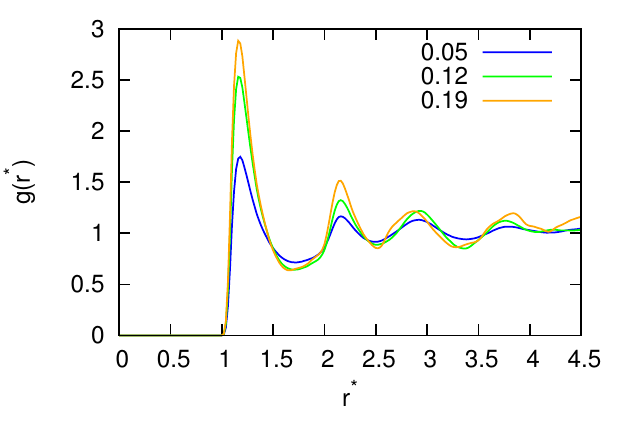}
\includegraphics[width=65mm]{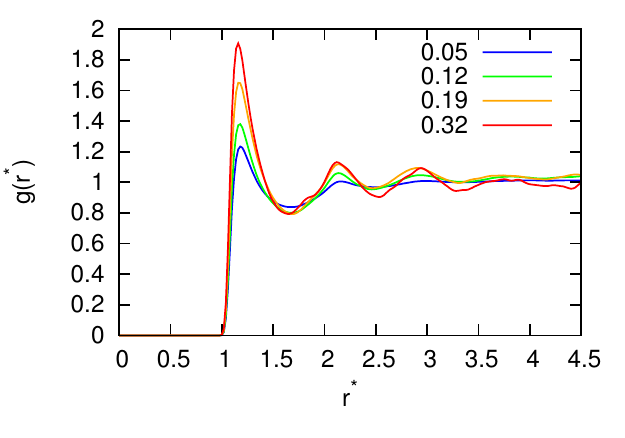}\\
  \caption{(Colour online) Solute-water pair correlation functions for various pressures for  sizes of the solution (a) $\sigma_{\rm{LJ}}=0.3$, (b)  $\sigma_{\rm{LJ}}=0.7$, and (c) $\sigma_{\rm{LJ}}=1.5$ for various pressures for depth of solute-solute LJ potential equal to $\varepsilon_{\rm{LJ}}=0.1$ for high temperature $T^*=0.24$ (right-hand) and for low temperature $T^*=0.18$ (left-hand). }
  \label{fig:sw}
\end{center}
\end{figure}
\begin{figure}[!htb]
	\begin{center}
(a)\includegraphics[width=65mm]{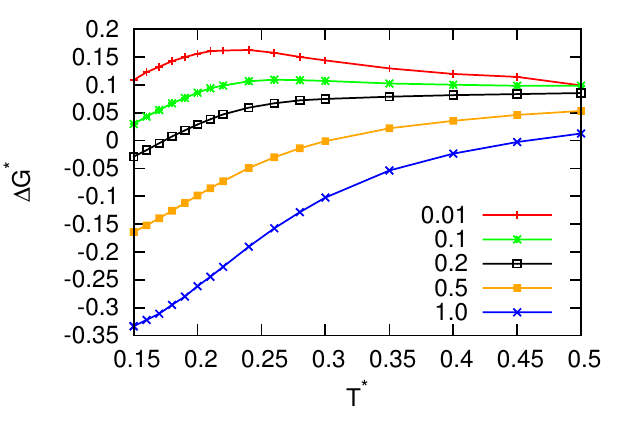}
\includegraphics[width=65mm]{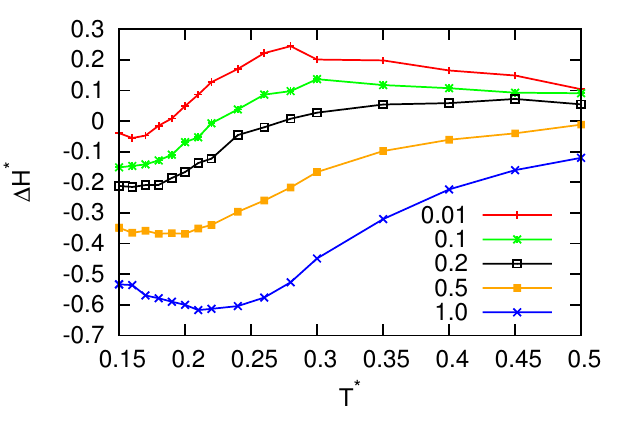}\\
(b)\includegraphics[width=65mm]{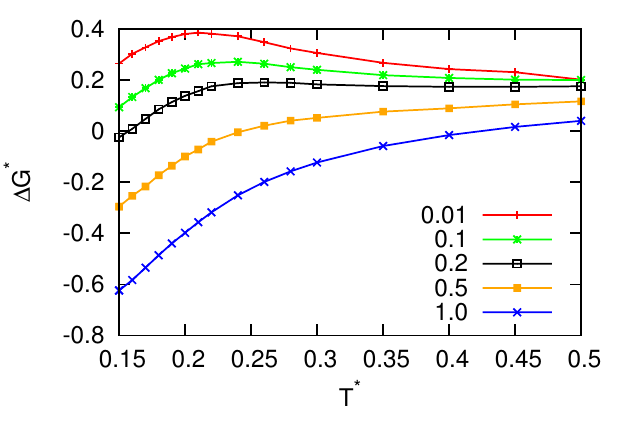}
\includegraphics[width=65mm]{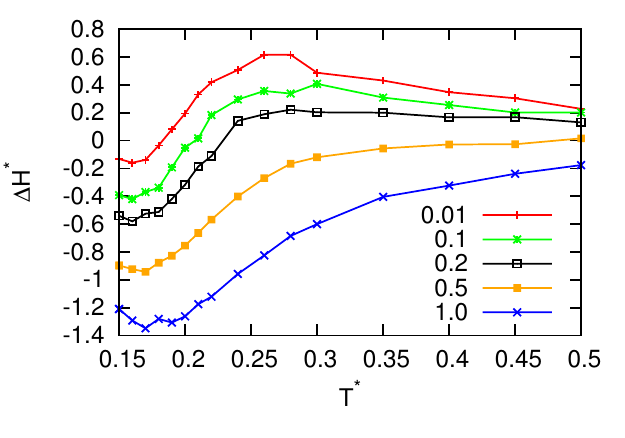}\\
(c)\includegraphics[width=65mm]{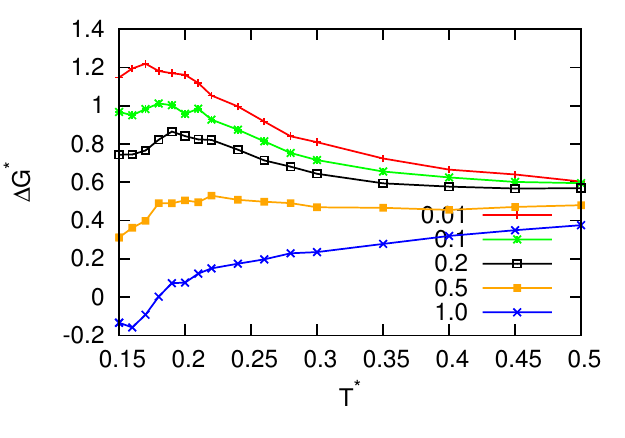}
\includegraphics[width=65mm]{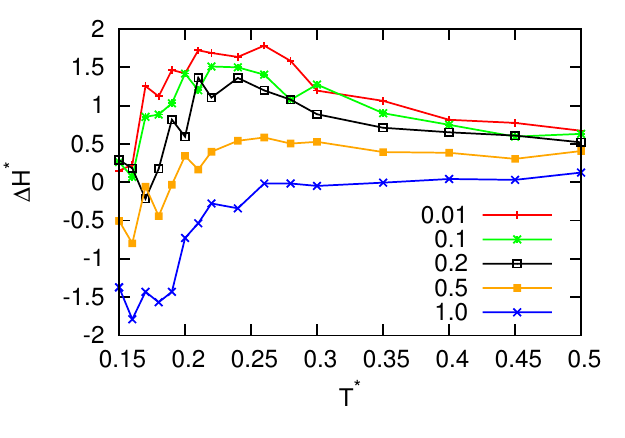}\\
  \caption{(Colour online) Temperature dependence of transfer free energy and transfer enthalpy of nonpolar solute for different depths of solute-solute LJ potential $\varepsilon_{\rm{LJ}}$ for  sizes of the solution (a)~$\sigma_{\rm{LJ}}=0.3$, (b)~$\sigma_{\rm{LJ}}=0.7$, and (c)~$\sigma_{\rm{LJ}}=1.5$ for pressure $p^*=0.19$.}
  \label{fig:epslj}
\end{center}
\end{figure}

In order to determine the solvation of nonpolar solute in 2D MB model of water we performed Monte Carlo computer simulations in the isothermal-isobaric ensemble with constant $N$, $p$, $T$~\cite{hansen,frenkel}. In order to mimic an infinite size of system of particles we used the periodic boundary conditions and the minimum image convention. All starting configurations were selected at random. In each move, we randomly tried to translate or rotate the random particle  or to change the volume. Probabilities for translation and rotation, change of volume and exchange of particles were the same. In one cycle we tried to translate and rotate each particle once on average and one change of volume per move of particles. The simulations were first equilibrated for as many cycles until equilibration was reached. The number of cycles depended on pressure and temperature. After equilibration, production runs were taken for minimum 20 series consistent with as many cycles as required to obtain well converged results. In the system we had from 100 to 500 particles depending on the density of the system. Thermodynamic quantities such as energy were calculated as statistical averages over the course of the simulations~\cite{frenkel}. Cut off of the potential was half-length of the simulation box. Increasing the number of particles had no significant effect on the calculated quantities. The cluster and bond analyses use an energy criteria wherein water molecules are considered bonded when their hydrogen bonded energy is less than $-0.05$. Small variations of this energy cutoff did not account for significant differences in the bonding state of the system. The mechanical properties such as enthalpy and volume were calculated as the statistical averages of these quantities over the course of the simulations~\cite{frenkel}. The heat capacity, $C_p$, the isothermal compressibility, $\kappa$, and the thermal expansion coefficient,
$\alpha$, are computed from the fluctuations~\cite{mbs}. The thermodynamic properties of a solvation process were computed using the Widom's particle insertion method~\cite{widom}. At each cycle, a ghost solute particle is randomly placed into the box and is not allowed to interact with the solvent particles. We then calculate the hypothetical interaction of the ghost particle with the solvent, and its weighted average. Thermodynamic quantities describing the solvation (free energy $\Delta G$, enthalpy $\Delta H$, entropy $\Delta S$ and heat capacity $\Delta c_p$) and the change in volume ($\Delta V$) follow from~\cite{mbs}:

\begin{eqnarray}
\Delta G&=&-\beta^{-1}\ln\left[\left\langle V\right\rangle_N^{-1}\left\langle V\exp\left(-\beta U\right)\right\rangle_N\right], \\
\Delta H&=&\frac{\left\langle H_{N+1}\exp\left(-\beta U\right)\right\rangle_N}{\left\langle\exp\left(-\beta U\right)\right\rangle_N}-\left\langle H_N \right\rangle_N, \\
T\Delta S&=&\Delta H - \Delta G, \\
k_bT^2\Delta c_p&=&\frac{\left\langle H_{N+1}^2\exp\left(-\beta U\right)\right\rangle_N}{\left\langle\exp\left(-\beta U\right)\right\rangle_N}-\frac{\left\langle H_{N+1}\exp\left(-\beta U\right)\right\rangle_N^2}{\left\langle\exp\left(-\beta U\right)\right\rangle^2_N}-\left\langle H_N^2\right\rangle_N+\left\langle H_N\right\rangle_N^2, \\
\Delta V &=&\frac{\left\langle V\exp\left(-\beta U\right)\right\rangle_N}{\left\langle\exp\left(-\beta U\right)\right\rangle_N}-\left\langle V\right\rangle_N,
\end{eqnarray}
where $\beta=\left(k_bT\right)^{-1}$. Angle brackets $<...>_N$ denote the averages throughout the run over all insertions. $U$ is the interaction of
ghost particle with the system and $H_{N+1}$ represents the enthalpy of the system with ghost particles. $\left\langle H_N\right\rangle_N$ and $\left\langle H_N^2\right\rangle_N$ stand for the average enthalpy and the average of squared enthalpy, respectively, of the system without the ghost particle.

\section{Results}

All the results are given in reduced units; the excess internal energy and temperature are normalized to the HB interaction parameter $\varepsilon_{\rm{HB}}$ ($E^*={E}/{\varepsilon_{\rm{HB}}}$, $T^*={k_{\rm{B}}T/{\varepsilon_{\rm{HB}}}}$) and all the distances are scaled to the HB characteristic length $r_{\rm{HB}}$ ($r^*={r}/{r_{\rm{HB}}}$).

In figure~\ref{fig:tt}, we calculated the temperature dependence of the density, $\rho^*$, the isothermal compressibility, $\kappa^*$, the thermal expansion coefficient, $\alpha^*$, the heat capacity, $c_p^*$ for bulk water for different pressures. Since $\kappa^*$ behaves as $1/p^*$ at high temperatures, we normalized the curves by multiplying them with $p^*$ so that all have the same high temperature limit. The lowest pressure $0.01$ is at below gas-liquid critical pressure, so we are in the region of phase transition which we can see as a jump for all 4 quantities. All other pressures are above the critical point, so we can see only the maxima in curves. These maxima in all three quantities are moving to higher temperatures as we increase pressure. The highest pressure $0.32$ is already so high that we no longer have density anomaly present for this model. The model has an area of anomalous behaviour where the density has maxima similar to real water for different pressures~\cite{water}. We can observe the density anomaly for pressures between $p^*=0.01$ and $p^*=0.3$ for this model. Here, we can also observe negative thermal expansion coefficients. In figure~\ref{fig:pp}, we calculated the pressure dependence of the same quantities ($\rho^*$, $\kappa^*$, $\alpha^*$, $c_p^*$) for bulk water for different temperatures in the region of density anomaly and outside the region. For density dependence we can see crossover of the curves in the region of density anomaly. We can also observe negative thermal expansion coefficients for these points. There are no much data available for pressure dependence in literature. We did some analytical modelling and we can observe the same trends in MB model as in analytical modelling of experimental water~\cite{urbic_jacs}. As last, we also checked the pair correlations function between two water molecules plotted in figure~\ref{fig:ww}. We presented the behaviour for high and low temperature for various pressures. At both temperatures we observe an increase of the pressure lower peak at distance 1 (HB peak). We can explain this as melting of hydrogen bonded structures. At the same time, we can notice that the peak at 0.7 is increasing (Lennard-Jones contact peak). At a very high pressure 1.0 both peaks have about the same size and the long range order is different, so we think that we have a different kind of liquid in comparison with lower pressures. 

Next, we studied the solvation of nonpolar solutes in this MB model of water. First we calculated the transfer free enthalpy and transfer enthalpy of nonpolar solute modelled as Lennard-Jones disk of various sizes. We first check the size dependence for isotherms at different pressures (figure~\ref{fig:tp2}) and isobars at different temperatures (figure~\ref{fig:tp3}). We can see the following form the curves: (1) bigger solutes are easier to inset at higher temperatures, position crossover depends on the temperature and pressure; (2) smaller solutes have linear dependence of the free enthalpy (up to size 1.0), bigger solutes have quadratic dependence; (3) transfer enthalpy has the lowest value for solutes of the size 1.0 (size of MB water, $r_{\rm{HB}}=1.0$; and (4) for high pressures ($p^*=1.0$) there is only one regime, only quadratic behaviour of transfer free enthalpy; transfer enthalpy also does not exhibit minima for reasons mentioned in the results of pure MB water. We can see that size dependence of solvation is similar to real 3D water. Bigger solutes are more solvable at high temperatures~\cite{dillreviewn}.
  
Next, we selected three different sizes of solutes from different regions; solutes smaller than water particles $\sigma_{\rm{LJ}}=0.3$ from the region of linear dependence of transfer free enthalpy, solutes of the same size as a water particles $\sigma_{\rm{LJ}}=0.7$ (this is the region of change from linear to quadratic behaviour of transfer free enthalpy), and solutes bigger than water molecules $\sigma_{\rm{LJ}}=1.5$ from region of quadratic behaviour. We checked how pressure affects the transfer quantities for the same type of solutes studied earlier~\cite{mbs}. The temperature dependence of the transfer quantities for different isobars is presented in figure~\ref{fig:tp1}. For all pressures, the transfer free energy is positive and we have the limit at high temperature to the value that it is proportional to the product of the pressure and volume of the solute. The dependence has maxima at a smaller pressure which is more pronounced at lower pressures. The position of the maxima depends on the solute sizes and shifts to lower temperatures as the size of the solute increases. The transfer enthalpy is negative and increases up to the maxima which is positive and then goes toward a limiting value which depends on pressure. Each curve has temperature at which the transfer enthalpy changes its sign. This temperature increases with an increase of the pressure and decreases with an increase of the size of the solute. These quantities behave differently at higher pressures. The transfer free enthalpy no longer has maxima. For smaller sizes, it increases toward the limit, for bigger sizes it decreases toward the limit. The behaviour for the same size as water has minima at temperature $T^*=0.18$ which is temperature where there is anomaly present at lower pressures. The transfer enthalpy is positive for all temperatures for higher pressures. The reason is that at these pressures the liquid is very dense and there are no cavities present for insertion of nonpolar solute. Thus, in order to insert the solute, we first need to spend energy to make a cavity and then the interaction energy between nonpolar solute and water is released, but all in all the change is positive. Next, we also checked solute-water pair correlation functions for different pressures (in figure~\ref{fig:sw}) for high and low temperature. For big solute and high pressure, the statistics were not good, so data are not present. The reason is high density and insufficiently big cavities to get good statistics. We can see that with an increase of the pressure, the contact peak increases, which is more pronounced at higher temperatures.  

Finally, we also checked how the depth of LJ potential between solutes changes the thermodynamics of solvation. In all previous studies, the value of this parameter was always kept equal to the value of MB. In figure~\ref{fig:epslj} we plotted the temperature dependence of transfer free energy, $\Delta G^*$, and transfer enthalpy, $\Delta H^*$, for various depths at a standard pressure of MB water model $p^*=0.19$. For small values of parameter $\varepsilon_{\rm{LJ}}$, the transfer free energy is positive and has maxima. At maxima, the solubility is the lowest. With an increase of the depth, the maxima first move to higher temperatures and later disappear and the values become negative. The transfer free enthalpy also decreases which means that solubility of the solute increases. We can see the same trend for all sizes, only that for bigger solutes it shifts to bigger depths. The transfer enthalpy has similar characteristics, but the shape is different. For smaller solutes, the transfer enthalpy is negative for all depths at low temperatures and increases with an increase of the temperature up to maxima and then tends toward the limiting value at high temperatures. This limiting value depends on the size of the solute. With an increase of the depth, the maxima disappear, but they are more pronounced for bigger solutes and shift toward lower temperatures. The maxima disappear at higher depths. 

\section{Conclusion}

In this work we have studied the aqueous solvation of a nonpolar solute. We used 2D MB model for water and LJ disks as nonpolar solutes. We studied the structure and thermodynamic properties  as a function of radius of nonpolar solute and solute-solute attraction, temperature and pressure in NPT Monte Carlo simulations. At low pressures, the model shows two different mechanisms, one for the solvation of large nonpolar solutes bigger than water and the second for smaller solutes. At higher pressures, we have only one kind of mechanism. The change of Lennard-Jones depth between solutes also leads to a stronger attraction with water and a negative transfer of free enthalpy and enthalpy.

\section{Acknowledgements}

The financial support of the Slovenian Research Agency through Grant P1-0201 as well as to projects J7-1816, J1-1708, N1-0186 and N2-0067 is acknowledged as well as National Institutes for Health RM1 award  RM1GM135136.

\ukrainianpart

\title{Залежність сольватації від тиску для неполярних розчинених речовин у простих моделях води}
\author{Т. Урбіч}
\address{Факультет хімії та хімічних технологій університету Любляни, вул. Вечна 113, SI-1000, Любляна, Словенія}

\makeukrtitle

\begin{abstract}
	У статті змодельовано водну сольватацію неполярної розчиненої речовини як функцію радіусу частинок, температури та тиску. Для моделювання Монте Карло в NPT-ансамблі використовується проста двовимірна т. зв. ``Мерседес-Бенц'' модель води. Раніше було показано, що ця модель здатна якісно передбачати об'ємні аномалії чистої води, а також вільну енергію, ентальпію, ентропію, теплоємність та зміни об'єму при введенні у воду неполярної розчиненої речовини. У даній роботі проведено більш детальне вивчення сольватації неполярних розчинених речовин з метою перевірки залежності цього процесу від тиску та ширшого діапазону температур і розмірів частинок. Дана модель демонструє два різні механізми: один --- для сольватації великих неполярних речовин, розміри частинок яких більші, ніж у води, та другий --- для менших частинок.
	\keywords Монте Карло, сольватація, неполярні розчинені речовини
\end{abstract}
\end{document}